\documentclass[conference]{IEEEtran}
\IEEEoverridecommandlockouts

\usepackage{cite}
\usepackage{amsmath,amssymb,amsfonts}
\usepackage{algorithmic}
\usepackage{graphicx}
\usepackage{textcomp}
\usepackage{multirow}
\usepackage[backref]{hyperref}

\usepackage{xcolor}

\def\BibTeX{{\rm B\kern-.05em{\sc i\kern-.025em b}\kern-.08em
    T\kern-.1667em\lower.7ex\hbox{E}\kern-.125emX}}
\begin{document}

\title{Optimizing Investment Strategies with Lazy Factor and Probability Weighting: A Price Portfolio Forecasting and Mean-Variance Model with Transaction Costs Approach\\
\thanks{*Corresponding author}}

\author{\IEEEauthorblockN{1\textsuperscript{st} Shuo Han}
\IEEEauthorblockA{\textit{College of Computer and} \\ \textit{Information Science} \\
\textit{Southwest University}\\
Chongqing, China \\
shuohan@email.swu.edu.cn}
\and
\IEEEauthorblockN{2\textsuperscript{rd} Yinan Chen}
\IEEEauthorblockA{\textit{College of Computer and} \\ \textit{Information Science} \\
\textit{Southwest University}\\
Chongqing, China \\
cyn54221467@email.swu.edu.cn}
\and

\IEEEauthorblockN{3\textsuperscript{th} Jiacheng Liu*}
\IEEEauthorblockA{\textit{College of Computer and} \\ \textit{Information Science} \\
\textit{Southwest University}\\
Chongqing, China \\
yc2020@email.swu.edu.cn}}




\maketitle

\begin{abstract}
Market traders often engage in the frequent transaction of volatile assets to optimize their total return. In this study, we introduce a novel investment strategy model, anchored on the 'lazy factor.' Our approach bifurcates into a Price Portfolio Forecasting Model and a Mean-Variance Model with Transaction Costs, utilizing probability weights as the coefficients of laziness factors. The Price Portfolio Forecasting Model, leveraging the EXPMA Mean Method, plots the long-term price trend line and forecasts future price movements, incorporating the tangent slope and rate of change. For short-term investments, we apply the ARIMA Model to predict ensuing prices.
The Mean-Variance Model with Transaction Costs employs the Monte Carlo Method to formulate the feasible region. To strike an optimal balance between risk and return, equal probability weights are incorporated as coefficients of the laziness factor.
To assess the efficacy of this combined strategy, we executed extensive experiments on a specified dataset. Our findings underscore the model's adaptability and generalizability, indicating its potential to transform investment strategies.
\end{abstract}

\begin{IEEEkeywords}
EXPMA, ARIMA, Quantitative investment, Markowitz Mean-Variance Model, Laziness Factor
\end{IEEEkeywords}

\section{Introduction}  
	
	
In recent years, there has been a notable transition in the investment sector, moving from primarily qualitative decision-making to a more data-centric, quantitative approach, driven by advancements in computer technology \cite{pan2021roles}. This paradigm shift, known as "Quantitative Investment," is fundamentally reshaping investment practices.

Quantitative investment capitalizes on comprehensive data analysis and complex modeling to identify promising investment opportunities and devise efficient strategies \cite{ashta2021artificial}. This systematic method effectively mitigates emotional reactions to temporary market volatilities, fostering long-term investor commitment. By harnessing positive market trends, it generates long-term returns and promotes market stability. The growing acceptance of this approach \cite{xiao2020stock} signals a promising future for quantitative investment in the financial sector. Alongside its expansion, there arises a need for more refined and sophisticated quantitative models to meet the dynamic market demands \cite{yi2022multifactor}. This ongoing requirement underscores the investment market's dynamism and the relentless search for innovative solutions \cite{alberg2017improving} to optimize investment strategies and results.

Despite their noted benefits, current quantitative investment methods often exhibit limitations in adaptability and generalizability \cite{leung2015validity,queiros2017strengths}. These approaches primarily rely on complex models to predict market trends, often overlooking inherent market variability \cite{petropoulos2022forecasting}. Moreover, such strategies usually adhere strictly to either long-term or short-term forecasting, neglecting the potential for synergistic insights offered by an integrated perspective \cite{ta2020portfolio}.

In response, this study introduces a new investment strategy model, designed for market traders frequently involved in volatile asset transactions to optimize their total return. Our proposed model, anchored on the 'lazy factor', consists of two main components: a Price Portfolio Forecasting Model and a Mean-Variance Model with Transaction Costs. The model utilizes probability weights as the laziness factor's coefficients.

The Price Portfolio Forecasting Model leverages the EXPMA Mean Method to establish the long-term price trend line, and forecasts price movements based on the tangent slope and rate of change. For short-term investments, we employ the ARIMA model to predict upcoming price trends. Simultaneously, the Mean-Variance Model, incorporating Transaction Costs, uses the Monte Carlo Method to define the feasible region. The model balances risk and return optimally by integrating equal probability weights as the laziness factor's coefficients.




\section{Related Work}

In this section, we provide a comprehensive overview of the current literature related to price-based portfolio investment strategies, encompassing both technical analysis and trading strategies.

\subsection{Investment methods}

Portfolio investment strategies use historical price or volume data to predict future trends by analyzing the embedded quantitative relationships \cite{binoy2022financial}. Traditional financial time series models, such as ARIMA \cite{zhang2022research,sun2022research,han2022quantitative}, gray forecasting models \cite{norouzi2020black}, and Markov chain Monte Carlo methods \cite{de2020using}, have shown excellent performance for short-term forecasting.


As machine learning continues to evolve, it offers investors the capability to decipher concealed patterns in asset prices, correlations, and volatilities. This enables the formation of portfolios optimized for risk-adjusted returns \cite{li2023optimization}. Buczynski et al. \cite{buczynski2021review} deliver a comprehensive review of numerous academic studies, providing an investment management viewpoint. Zhang \cite{zhang2020research} advocates for the use of machine learning algorithms in stock selection, highlighting their effectiveness in formulating investment strategies. In a parallel vein, Yi \cite{yi2022multifactor} presents a multifactor investment methodology underpinned by pivotal elements of corporate finance and valuation, further streamlining portfolio selection through a machine learning-based classification system.

\subsection{Trading strategies }
Trading strategies aim to guide investors in wisely allocating their capital, managing risk-reward balance, and achieving wealth accumulation. The development of capital markets and the rapid growth of investment products led to the introduction of portfolio investment strategies \cite{cesari2013trading}.

Markowitz and Todd \cite{markowitz2000mean} first proposed the theory of portfolio selection. The mean-variance analysis, a key portfolio methodology, is continually optimized and adapted. However, actual results often diverge from predictions. Jobson and Korkie \cite{jobson1980estimation} suggested that complex investment strategies may not perform as well as simpler ones due to estimation errors. Additionally, equal-weighted investment strategies are also gaining popularity.

\section{Our Approach}  

This section details our proposed Price Portfolio-Forecasting Model, designed to predict future price trends for optimal portfolio strategizing. By integrating technical analysis and trading-based strategies, we have developed a comprehensive model. This leverages the Exponential Moving Average (EXPMA) Mean Method for long-term price trend analysis and employs the AutoRegressive Integrated Moving Average (ARIMA) model for short-term price predictions.

\subsection{Price Portfolio-Forecasting Model}
The insights derived from previous research have served as an inspiration for our approach. As a case study, we consider a specific day and engage in information extraction from the historical prices of gold and Bitcoin. The goal is to anticipate future price trends, which then serves as a determinant in devising an optimal portfolio strategy.

\subsubsection{Model for Long-Term Forecasting based on EXOMA Averages}
Drawing parallels with the stock market, we initially employ the Moving Average (MA) and Simple Moving Average (SMA) methods to compute the 5-day, 20-day, and 60-day averages of historical prices. These averages over distinct time intervals are then connected to shape short-term, medium-term, and long-term trend lines, respectively. The calculation formula is expressed as follows:


\begin{equation}
\begin{aligned}
MA(N)=\frac{1}{N} \sum_{i=1}^{N} x_{i j} \quad
\end{aligned}
\end{equation}

Here, $j=1,2$ with $1$ representing gold and $2$ denoting bitcoin.
     
We improve the model by using exponential and smoothing coefficients to obtain the EXPMA model considering the lag of the mean in tracking the price makes it does not reflect the trend change in time. The calculation formula is as follows.  
    	

\begin{equation}
\begin{aligned}
\operatorname{EMA}(N)&=\eta_{N}\left(x_{N}-\operatorname{EMA}(N-1)\right)+\operatorname{EMA}(N-1) \\
\eta_{N}&=\frac{\sum_{i=1}^{N} x_{ij}}{N}, \quad j=1,2
\end{aligned}
\end{equation}
    \subsubsection{Short-Term Forecasting based on ARIMA Series}
    	From the previous analysis, the results of short-term forecasting are closer to the real price. Thus, the ARIMA (autoregressive moving average) model is considered to forecast the short-term prices to obtain the possible prices for the next day and provide the expected return value for the investment strategy of that day.
    	
    	In the ARIMA model, we use the observed values of the past price series of gold to derive future prices. The future prices are expressed as linear functions of the current and lagged periods of the lagged and stochastic disturbance terms. The general form of the model is as follows:
    	
    	\begin{equation}
    		\begin{array}
    			{c}	Y_{t}=c+\alpha_{1} Y_{t-1}+\cdots+\alpha_{p} Y_{t-p}+\epsilon_{t}+ \\   \beta_{1} \epsilon_{t-1}+\cdots+\beta_{q} \epsilon_{t-q}
    			
    		\end{array}
    	\end{equation}
    
    The expansion of the time interval will increase the error of the price forecast. We keep adjusting the length of the interval to find the optimal value in the process of forecasting. Eventually, we found that the traditional forecasting model generally considers the amount of data within 100 for data with large price changes such as gold and bitcoin combined with the information. Too much data, on the contrary, can lead to prediction results deviating from expectations.
  
\subsection{Integration of Transaction Costs into the Price Portfolio-Forecasting Model using the Mean-Variance Model}
\subsubsection{Gold trading days}

Markowitz treats return as a random variable, takes the mean of asset return as the expected return  \cite{lin2022improved}, and defines risk as the coefficient of fluctuation size of return - variance, which is actually the degree of penalty for deviation of asset return from the expected return. With expected return maximization and risk minimization as the objective function, the problem is treated as a multi-objective planning problem \cite{xidonas2017robust}.

Since the daily [cash, gold, bitcoin] portfolio needs to be obtained, cash is also considered as an asset in this paper, so the model includes three risky assets: cash, gold, and bitcoin, but the complexity of the model is not increased too much due to the special nature of cash.

The rates of return for the three risky assets are represented by $R_{t i}$, and for cash, it is understood due to its unique nature. For the firm, the rate of return is defined as the ratio of net profit to the average capital employed. In this study, we estimate the rate of return based on the current asset prices and make a short-term forecast for tomorrow's prices.

\begin{equation}
	\begin{array}
		{c} R_{t i}=\ln \left(y_{(t+1) i} / x_{t i}\right), \quad i=1,2
	\end{array}
\end{equation}

The weights of cash, gold and bitcoin obtained under the investment strategy of date t (i.e., the proportion of actual holdings) are $\omega_{t0}$,$\omega_{t1}  $ and $\omega_{t2}$, 
 $\omega_{t0}+\omega_{t1}+\omega_{t2}=1$.

Then the expected return rate is the portfolio under this strategy is the weighted average of the expected returns of gold and bitcoin, denoted as $R_t=R_{t1 }\omega_{t1}+R_{t2} \omega_{t2}$.

The risk of the portfolio is expressed as:

\begin{equation}
\begin{array}{c}
	\sigma_{t}^{2}=\sum_{i=0}^{2} \sum_{j=0}^{2} \omega_{t i} \omega_{t j} \operatorname{Cov}_{t j j} 
    = \omega_{t 0}^{2}  \sigma_{t 0}^{2} \\+ \omega_{t 1}^{2} \sigma_{t 1}^{2}+   \omega_{t 2}^{2} \sigma_{t 2}^{2}+   \omega_{t 0} \omega_{t 1} \rho_{t 01} \sigma_{t 0}^{2}  \sigma_{t 1}^{2} \\ + \omega_{t 0} \omega_{t 2} \rho_{t 02} \sigma_{t 0}^{2} \sigma_{t 2}^{2}+\omega_{t 1} \omega_{t 2} \rho_{t 12} \sigma_{t 1}^{2} \sigma_{t 2}^{2}
\end{array}
\end{equation}

The notation $\mathrm{i} \neq \mathrm{j}$ signifies that $\operatorname{Cov}{i j}$ represents the covariance of the returns of gold and bitcoin (and vice versa), which is the product of the correlation coefficient of their returns, $\rho{t i j}$, and the standard deviations of their individual returns, $\sigma_{t i}$ and $\sigma_{t j}$. This is mathematically represented as $\operatorname{Cov}{t i j}=\rho{t i j}\sigma_{t i}\sigma_{t j}$. The portfolio risk escalates progressively as the correlation coefficient varies between -1 and 1. Unless the correlation coefficient equals 1, the risk of a binary portfolio, $\sigma_{t}$, consistently remains below the weighted average of the separate investment risks. This demonstrates that investment in a portfolio can mitigate investment risk.

Assuming that the initial assets are $ W_{0}$, the total value of assets after the investment strategy at date t:
\begin{equation}
	\begin{array}{c}
	W_{t}=W_{t-1}\left(1+R_{t}\right) \\
	-\alpha_{1} x_{t 1}\left|\frac{W_{t-1} \omega_{(t-1) 1}}{x_{(t-1) 1}}-\frac{W_{t-1}\left(1+R_{t}\right) \omega_{t 1}}{x_{t 1}}\right| \\ 
 \\
 -  \alpha_{2} x_{t 2}\left|\frac{W_{t-1} \omega_{(t-1) 2}}{x_{(t-1) 2}}-\frac{W_{t-1}\left(1+R_{t}\right) \omega_{t 2}}{x_{t 2}}\right|
	\end{array}
\end{equation}

The term $\frac{W_{t-1} \omega_{(t-1) 1}}{x_{(t-1) 1}}$ represents the gold holdings on the day before trading day t, and $\frac{W_{t-1}\left(1+R_{t}\right) \omega_{t 1}}{x_{t 1}}$ represents gold holdings after the investment decision on trading day t. The absolute difference between the two symbolizes the trading volume of gold on day t. A similar interpretation applies to bitcoin. (The ultimate goal is to obtain the decision after W on 2021.2.2.)

The investor aims for the highest possible return rate and the lowest possible uncertainty risk \cite{cheng2019risk}. The ideal objective is to balance these two conflicting goals. The conventional Markowitz mean-variance model pursues maximum expected return and minimum risk. Considering transaction costs that accompany every gold or bitcoin purchase or sale, this paper proposes the maximum total asset value on transaction day t as the objective function.
\\
\begin{equation}
\begin{array}{c}
\left\{\begin{array}{l}
\text { Min } \sigma_{t} \\
\text { Max } W_{t}
\end{array}\right.
\end{array}
\end{equation}
\begin{equation}
\text{s.t.}\quad \sum_{i=0}^{2} \omega_{ti} = 1
\end{equation}
\begin{equation}
	\begin{array}{c}
	0 \le \omega_{t i} \leq 1, i=0,1,2
	\end{array}
\end{equation}
\\
\subsubsection{Non-Trading Days for Gold (Weekends and Holidays)}
On non-trading days for gold, the gold holdings and closing prices are carried over from the previous trading day. The model then treats only cash and bitcoin as risky assets, with the total asset value equating to the previous day's total value of cash and bitcoin. We adapt our model for these two risky assets, which can be calculated similarly.

\subsection{Portfolio Investment Strategy Considering the Laziness Factor}
The mean-variance approach used practically, necessitates precise estimates of individual asset returns' mean and variance. Estimation errors can cause the mean-variance efficient portfolio to diverge from the true efficient portfolio. The equal-weight investment strategy, on the other hand, eliminates estimation errors by assigning equal weights to portfolio assets without requiring parameter estimation or optimization techniques. Additionally, this strategy is unaffected by external or personal investor factors \cite{haensly2020risk}.

Hence, we propose a blend of the equal-weight strategy and a mean-variance model incorporating transaction costs. This includes a quadratic combination of gold and bitcoin weights for different risk levels, thereby considering the laziness factor. The approach aims to stabilize weights and reduce local optimal solution responsiveness. 

\begin{itemize}
    \item Equal weights
    
    Cash, gold, and bitcoin within the portfolio are given equal weights:
$ \overline{\omega_{t 0}}=\overline{\omega_{t 1}}=\overline{\omega_{t 2}}=1 / 3 $. These weights are irrespective of asset return distribution characteristics.

\item The laziness factors

The laziness factors, $ \delta_{1}$ and $ \delta_{2}$, are considered. Here, $ \delta_{1}$ and $ \delta_{2}$ denote the portfolio coefficients of the equal-weighted strategy and the mean-variance model considering transaction costs, respectively.
Hence, $\left(\begin{array}{l}\omega_{t 0}{ }^{\prime} \ \omega_{t 1}{ }^{\prime} \ \omega_{t 2}{ }^{\prime}\end{array}\right)=\delta_{1}\left(\begin{array}{l}\overline{\omega_{t 0}} \ \overline{\omega_{t 1}} \ \overline{\omega_{t 2}}\end{array}\right)+\delta_{2}\left(\begin{array}{c}\omega_{t 0} \ \omega_{t 1} \ \omega_{t 2}\end{array}\right)$, with constraints $\delta_{1}+\delta_{2}=1,0 \leq \delta_{1}, \delta_{2} \leq 1$.

\item Weighting Coefficients Determination

The weights $ \left(\begin{array}{c}\omega_{t 0} \ \omega_{t 1} \ \omega_{t 2}\end{array}\right)$, derived from the model, function as ideal weights. These are combined with equal weights to temper potential discrepancies and significant fluctuations. Combination coefficients are computed considering the Relative Strength Index (RSI), rate of change ($\beta$), and mid to long-term trend indicators (slope of moving average $\gamma$).
\end{itemize}


The RSI, calculated as the average closing gain over the sum of average closing gain and loss, times 100\%, indicates trading tendencies and trend momentum, assisting short-term investments.

$\beta$ denotes the ratio of the price increase and decrease days for cash, gold, and bitcoin in the fortnight before the trading day, illustrating asset price uncertainty.


\begin{table}[h]
\centering
\begin{tabular}{c|l}
	\hline
	Impact Indicators & \multicolumn{1}{c}{Results}                                                                                                                 \\ \hline
	RSI               & \begin{tabular}[c]{@{}l@{}}Higher RSI signifies more substantial buying power.  \end{tabular}          \\ \hline
	$\beta $       & \begin{tabular}[c]{@{}l@{}}Higher $\beta $ indicates a greater number of price \\increase days in the period.\end{tabular}                                  \\ \hline
	$\gamma $   & \begin{tabular}[c]{@{}l@{}}Larger $\gamma $ implies a higher upward price movement \\following the trading day. \end{tabular} \\ \hline
\end{tabular}
\end{table}

The influence of all the above three indicators on the ideal weight is to make it higher, so this paper considers expressing the combination coefficient as:
\begin{equation}
	\begin{array}{c}
\delta_{1}=0.8 \frac{R S I}{R S I+\beta+\gamma}, \quad \delta_{2}=1-\delta_{1}
\end{array}
\end{equation}

 The combination coefficients were calculated based on the five years of data given in the title and the mean value was processed to obtain:  $\delta_{1}=0.7, \delta_{2}=0.3 $.

\subsection{Calculating the total asset value}
 Starting from 11/9/2016, since there is no historical data in the initial period to refer to help make decisions, this paper treats the previous month as an empty period to accumulate historical data and make portfolio decisions starting from 11/10/2016.

 After the above steps, we obtain the portfolio weights $ \left(\begin{array}{c}\omega_{t 0}{ }^{\prime} \\ \omega_{t 1} \text {, } \\ \omega_{t 2}{ }^{\prime}\end{array}\right)$  for the adjusted trading day t. Therefore, after the daily portfolio decision is completed, the total asset value for the day is: $\left[C_{t}, G_{t}, B_{t}\right]=\left[\omega_{t 0}^{\prime} W_{t}, \omega_{t 1}^{\prime} W_{t}, \omega_{t 2}^{\prime} W_{t}\right]$. And $ \left[C_{0}, G_{0}, B_{0}\right]=\left[\omega_{00}^{\prime} W_{0}, \omega_{01}^{\prime} W_{0,}, \omega_{02}^{\prime} W_{0}\right]$ , which $ W_{0}=1000,\left(\begin{array}{l}\omega_{00}^{\prime} \\ \omega_{01}^{\prime} \\ \omega_{02}^{\prime}\end{array}\right)=\left(\begin{array}{l}1 \\ 0 \\ 0\end{array}\right)$ .

\section{Experiments}
\subsection{Data Description}
	This study utilizes a portfolio consisting of gold and Bitcoin trades, incorporating historical gold and Bitcoin price data: London Bullion Market Association's daily gold prices (9/11/2016 - 9/10/2021) and NASDAQ's Bitcoin daily prices (9/11/2016 - 9/10/2021).
\subsection{Performance Evaluation}
\subsubsection{Long-Term Forecasting Performance}
	The 2021 Bitcoin price trend serves as an example, with results illustrated in the subsequent chart.



\begin{figure}[thb] \centering
    \includegraphics[width=0.23\textwidth]{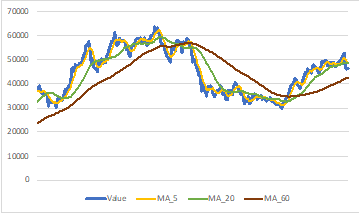}
    \includegraphics[width=0.23\textwidth]{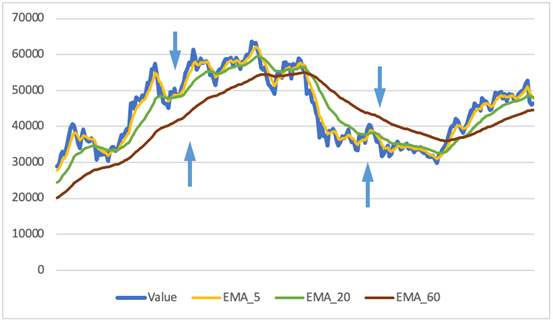}
        \makebox[0.235\textwidth]{\small (a) MA}
    \makebox[0.235\textwidth]{\small (b) EMA}
    \caption{Trend chart based on MA and EMA model.}
    \label{fig:figure2}
\end{figure}

	

Fig \ref{fig:figure2} shows that the EMA trend line is more accurate than MA in predicting price trends. Short-term price trends help predict next-day prices, while long-term trends provide insight into overall price movements over extended periods. The chart also identifies instances, marked by blue arrows, where short-term trends contradict long-term ones.

	\begin{figure}[htbp]
		\begin{minipage}[t]{0.5\linewidth}
			\centering
			\includegraphics[width=0.87\textwidth]{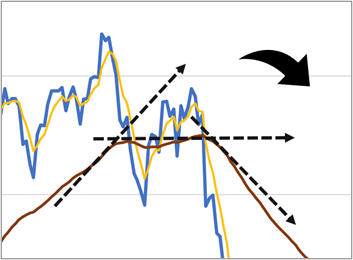}
			
		\end{minipage}%
		\begin{minipage}[t]{0.5\linewidth}
			\centering
			\includegraphics[width=0.87\textwidth]{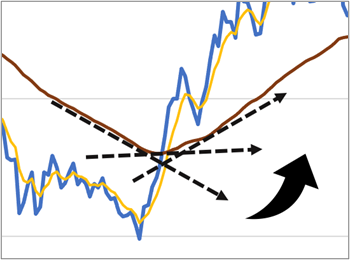}
			
		\end{minipage}
		\caption{Inflection point charts of the EMA long-term trend line (partial)}
  \label{Inflection}
	\end{figure}

As illustrated in Figure \ref{Inflection}, two inflection points on the long-term trend line are significant. The change in slope K at these points indicates a sudden alteration in the asset price trend. Specifically, a convex point presages a future decrease in price, while a concave point forecasts a future increase \cite{warren2019choosing}. This transition is quantified by the rate of change in slope K’ between consecutive tangent lines.

Informed by these findings, we suggest the following strategy: The anticipation of a price rise materializes when short-term moving averages intersect and surpass the long-term averages. Conversely, a prediction of a price fall emerges when short-term moving averages intersect and fall beneath the long-term averages.

\paragraph{Smoothness Test}

For the ARIMA model to predict time series data effectively, stability is crucial to capture the regularity \cite{du2019implementation}. Given the volatility of gold prices, the ADF test is initially applied to check the unit root of the original series. The original non-stationary time series is then converted into a stationary one through differential transformation. The smoothness test results are presented in the ensuing table.


\begin{table}[h]
	\centering
	\caption{ADF Inspection Form}
	\begin{tabular}{c|ccc|ccl}
		\hline
		\multirow{2}{*}{\begin{tabular}[c]{@{}c@{}}Difference \\ Order\end{tabular}} & \multirow{2}{*}{t}         & \multirow{2}{*}{P}           & \multirow{2}{*}{AIC}        & \multicolumn{3}{c}{Threshold Value}                                               \\ \cline{5-7} 
		&                            &                              &                             & 1\%                       & 5\%                        & \multicolumn{1}{c}{10\%} \\ \hline
		0                                                                            & 0.321                      & 0.978                        & 530.564                     & -3.507                    & -2.895                     & -2.585                   \\
		1                                                                            & -5.988                     & 0.000***                     & 522.718                     & -3.507                    & -2.895                     & -2.585                   \\
		2                                                                            & -7.477                     & 0.000***                     & 528.297                     & -3.51                     & -2.896                     & -2.585                   \\ \hline
	\end{tabular}
\end{table}

	As can be seen from the above table, at a difference score of 0, the significance P-value is 0.978, indicating that the original gold price series is not smooth. At a difference score of 1, the significance P-value is 0.000***, which presents significance at the level, and the differenced price series to reach a smooth state. The smoothing results are as follows.

\begin{figure}[thb] \centering
    \includegraphics[width=0.23\textwidth]{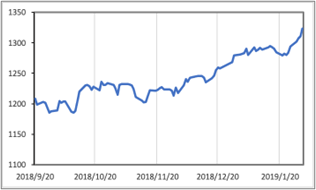}
    \includegraphics[width=0.23\textwidth]{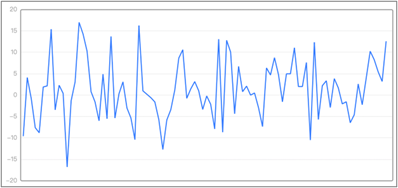}
        \makebox[0.235\textwidth]{\small (a) Original}
    \makebox[0.235\textwidth]{\small (b) Smoothed}
    \caption{Smoothness Test.} \label{fig:figure2}
\end{figure}
	
			
		

\paragraph{ARIMA Model}
	
The development of the ARIMA model commences with the establishment of the model order, leveraging autocorrelation coefficients, partial autocorrelation coefficients, and other parameters. An optimal model is identified in accordance with the Akaike information criterion for fixed-order selection.

The subsequent stages encompass parameter estimation and validation, including the significance of parameters, the robustness of the model, and an examination of whether the residual series constitutes a white noise series \cite{ariyo2014stock}. If the model does not pass the test, the selection of a new model becomes necessary. The process culminates with the presentation of correlation results.
	
	\begin{table}[h]
		\centering
		\caption{Model Narameter Test Table}
		\begin{tabular}{c|cc}
			\hline
			\multirow{2}{*}{Item}                                   & \multirow{2}{*}{Symbol} & \multirow{2}{*}{Value} \\
			&                         &                        \\ \hline
			\multicolumn{1}{c|}{\multirow{2}{*}{Number of Samples}} & Df Residuals            & 87                     \\
			\multicolumn{1}{c|}{}                                   & N                       & 93                     \\
			\multicolumn{1}{c|}
   {\multirow{5}{*}{Q Statistics}}      & Q6(p)                   & 0.039(0.844)           \\
			\multicolumn{1}{c|}{}                                   & Q12(p)                  & 6.124(0.409)           \\
			\multicolumn{1}{c|}{}                                   & Q18(p)                  & 8.165(0.772)           \\
			\multicolumn{1}{c|}{}                                   & Q24(p)                  & 13.321(0.772)          \\
			\multicolumn{1}{c|}{}                                   & Q30(p)                  & 27.206(0.295)          \\
			\multicolumn{1}{c|}{Information}                        & AIC                     & 619.465                \\
			\multicolumn{1}{c|}{Fitting Optimization}               & R\textasciicircum{}2    & 0.964                  \\ \hline
		\end{tabular}
	\end{table}

	From the above table, Q6 does not show significance at the level, and the gold price residual series is a white noise series. Meanwhile, the goodness-of-fit R-squared of the model is 0.964, which is a good fit.
	
	\paragraph{Future Price Forecast}
	
	The final results of the model solution are as follows.

	\begin{equation}
		\begin{array}
			{c}y_{(t)}=1.238+0.684 y_{(t-1)}-0.81 y_{(t-2)} \\ -0.67 \varepsilon_{(t-1)}+1.0 \varepsilon_{(t-2)}
		\end{array}
	\end{equation}

	\begin{figure}[h]
		\centering
		\includegraphics[width=0.38\textwidth]{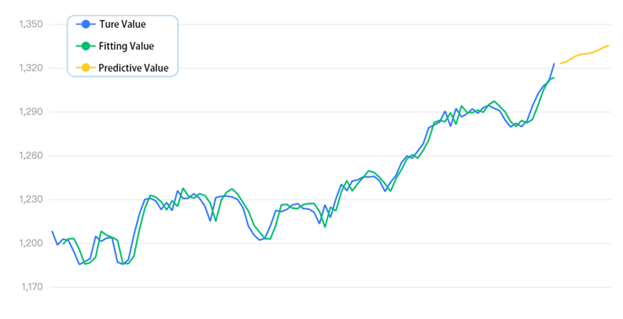}
		\caption{Figuer of the forecast results of the gold price series}
	\end{figure}

	The predicted price of gold for the next day is 1322.966 and the actual price is 1318.7, with an error rate of about 0.32$\%$, which shows the high prediction accuracy.

\subsubsection{Portfolio-Investment Strategy Model with Laziness Factor}

\begin{figure}[thb] \centering
    \includegraphics[width=0.23\textwidth]{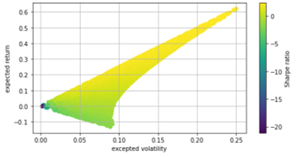}
    \includegraphics[width=0.23\textwidth]{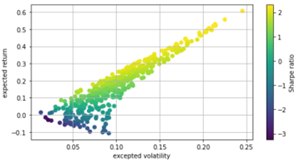} 
    \\
    \includegraphics[width=0.23\textwidth]{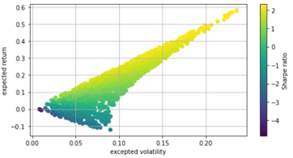}
    \caption{The Result of Monte Carlo method.} \label{fig:figure3}
\end{figure}

We define a feasible set for three risky assets using the complete range of weight ratios, based on two months of data. The Monte Carlo method generates data sets at 500, 1000, and 10,000 intervals. Results, including the risk and expected return, are visualized in Figure \ref{fig:figure3}. The minimum risk is at the apex of the left convex curve.

Figure \ref{fig:figure3} also displays cumulative returns for cash, gold, and bitcoin. Notably, bitcoin shows a consistent upward trend, outperforming in later stages, reflecting its increasing dollar value over the last five years.
	
			
			
			

Within the feasible set, two key portfolio characteristics can be discerned: 1) the portfolio that provides the maximum expected return for a specific level of risk; 2) the portfolio that maintains the least risk for a given expected return. These portfolios represent optimal points, and together they establish a curve referred to as the minimum variance frontier or the efficient frontier for risky assets.

		\begin{figure}[h]
			\centering
			\includegraphics[width=0.345\textwidth]{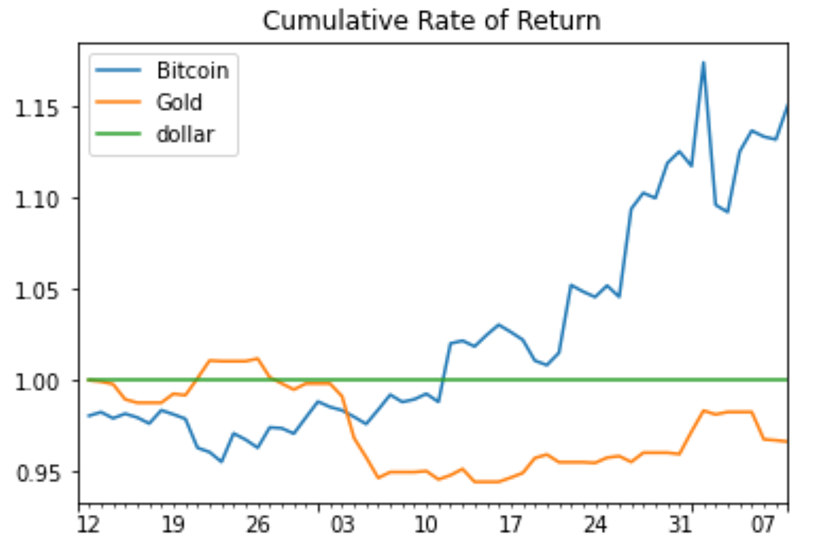}
			\caption{Cumulative Rate of Return}
		\end{figure} 

Further analysis allows for the identification of the portfolio weight on the efficient frontier where the Sharpe ratio reaches its maximum. This point represents the tangency between the capital allocation line (CAL) and the efficient frontier, known in the field as the Capital Market Line (CML).  

\begin{figure}[h]
\centering
\includegraphics[width=9.5cm,height=8cm]{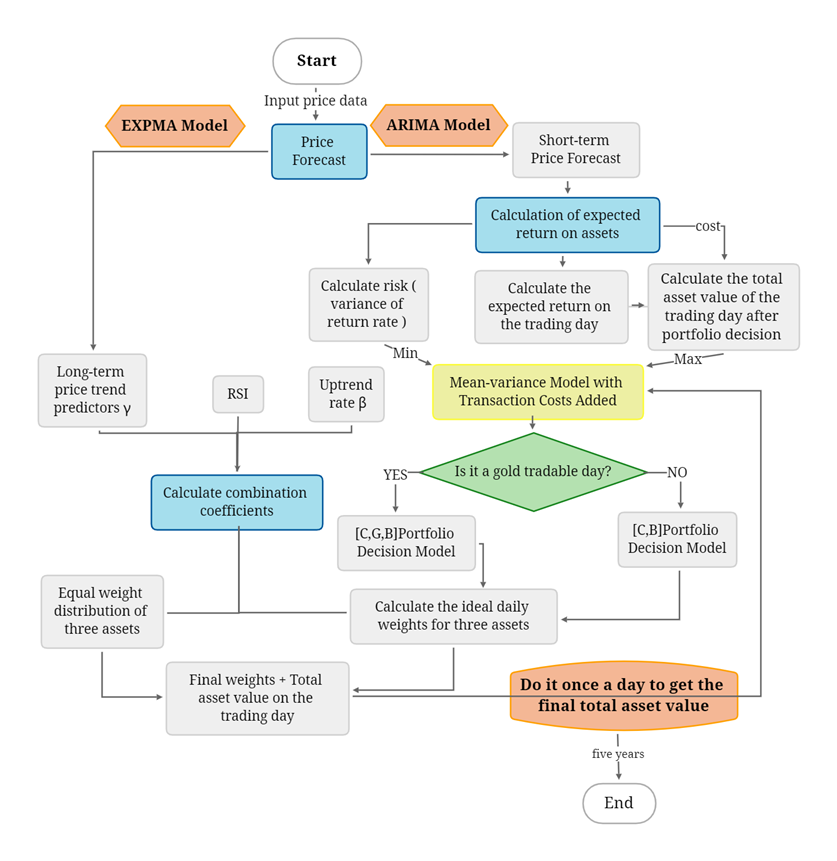}
\caption{ Flowchart for solving the optimal portfolio strategy}
\end{figure} 
		
		\begin{table}[]
				\centering
			\caption{Change in asset value over a five-year period}
			\begin{tabular}{ccccc}
				\hline
				Date                                              & C                                                 & G                                                 & B                                                 & Total Return                                      \\ \hline
				2016/9/11                                         & 1000                                              & 0                                                 & 0                                                 & 1000                                              \\
				\begin{tabular}[c]{@{}c@{}}.\\ .\\ .\end{tabular} & \begin{tabular}[c]{@{}c@{}}.\\ .\\ .\end{tabular} & \begin{tabular}[c]{@{}c@{}}.\\ .\\ .\end{tabular} & \begin{tabular}[c]{@{}c@{}}.\\ .\\ .\end{tabular} & \begin{tabular}[c]{@{}c@{}}.\\ .\\ .\end{tabular} \\
				2018/9/7                                          & 1808.8                                            & 0.74323                                           & 0.60436                                           & 6618.17                                           \\
				2018/9/8                                          & 1813.2                                            & 0.73423                                           & 0.61716                                           & 6622.37                                           \\
				2018/9/9                                          & 1817.6                                            & 0.73423                                           & 0.6283                                            & 6647.67                                           \\
				\begin{tabular}[c]{@{}c@{}}.\\ .\\ .\end{tabular} & \begin{tabular}[c]{@{}c@{}}.\\ .\\ .\end{tabular} & \begin{tabular}[c]{@{}c@{}}.\\ .\\ .\end{tabular} & \begin{tabular}[c]{@{}c@{}}.\\ .\\ .\end{tabular} & \begin{tabular}[c]{@{}c@{}}.\\ .\\ .\end{tabular} \\
				2021/9/7                                          & 25748                                             & 2.1982                                            & 2.0051                                            & 135339.5                                          \\
				2021/9/8                                          & 25811                                             & 2.2248                                            & 2.2666                                            & 135884.5                                          \\
				2021/9/9                                          & 25873                                             & 2.2274                                            & 2.3081                                            & 136216.1                                          \\ \hline
			\end{tabular}
		\end{table}
		
		Finally, the initial \$1000 is calculated in a loop to get the final total value of \$136216.1 on 10/9/2021. It shows that our model achieves a remarkable annualized rate of return of 27.04$\%$, which demonstrate the strong profitability.
		
	\begin{figure}[h]
		\centering
		
  \includegraphics[width=0.345\textwidth]{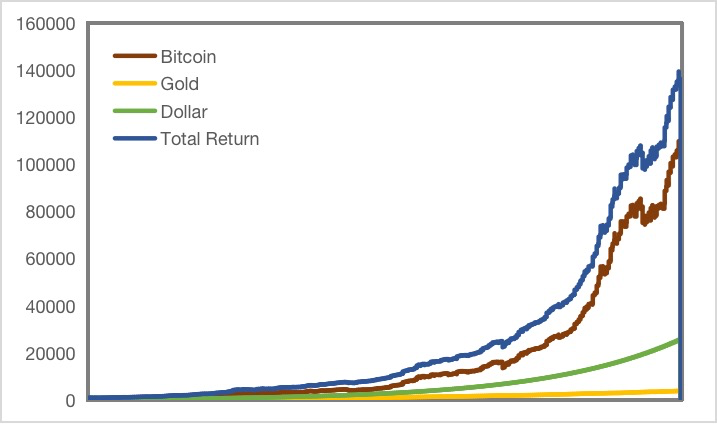}
		\caption{Figuer of the best daily portfolio}
\label{fig:fig81}
	\end{figure} 

    As can see from the trend in Fig \ref{fig:fig81}, Bitcoin has the largest impact on total returns, with gold only accounting for a small percentage.
    
\subsection{Experimental Evaluation of Portfolio Investment Strategy Model}
\subsubsection{Intra-Model Test}

\begin{figure}[h]
\centering
\includegraphics[width=0.345\textwidth]{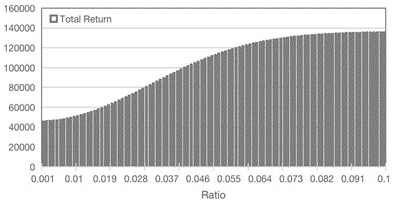}
\caption{Laziness factor against the total return benefit}
\label{fig:Laziness factor}
\end{figure}

The Markowitz mean-variance model brings substantial benefits, particularly diversification, a principle advocated by Nobel laureate economist James Tobin: "Don't put your eggs in one basket". This principle allows investors to reduce risk without sacrificing expected returns by combining different assets \cite{lefebvre2020mean}. A backtest conducted on trades using this model yielded a profitability rate of 135.2161\% and a maximum retracement rate of 2.5044\%. When combined with real market data, it shows enhanced returns and diminished risk, thus illustrating the efficacy of our developed portfolio optimization model. To temper the sensitivity of the ideal weight results to the local optimal solution, we included a laziness factor, combining equal probability weights (i.e., 1/3) with the ideal weights through a quadratic combination. Adjusting the combination coefficients before equal weights allows us to ascertain the final asset value as these coefficients change, with the results shown in Fig \ref{fig:Laziness factor}.

The total value of the final assets obtained when   tends to be optimal. It can be seen that the portfolio factors we have selected can achieve the optimal portfolio strategy.
		
		\subsubsection{Out-of-Model Tests}
		A control group is set up: no laziness factor is added, i.e. no quadratic combination of equal probability weights 1/3 with the ideal weights is added to stabilize the weights, and only the ideal weights derived from the mean-variance model with transaction costs are used for portfolio decisions.
		
\begin{figure}[h]
			\centering
			\includegraphics[width=0.345\textwidth]{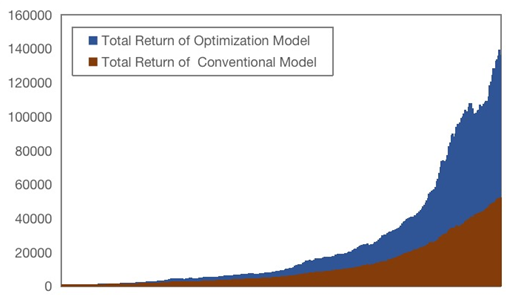}
			\caption{Chart comparing the total return of the traditional model with the total return of the
				optimized model
			}
   \label{figChart}
		\end{figure}

		The daily portfolio strategy and the total asset value for each day over the five years are derived and the results are compared with those of the optimization model. The results are shown in Fig \ref{figChart}.

		As can be seen from the results, the daily total asset value of our built model is generally higher than the results without the laziness factor, and the final 9/10/19 total value is higher than the results without the laziness factor by approximately 160.81$\%$. Thus, it can be shown that the optimization model we have built provides the best portfolio strategy.
		
		\section{Conclution}
		In this paper, we propose a dual-model portfolio investment strategy that amalgamates long-term (EXPMA-based) and short-term (ARIMA-based) price forecasting with an optimized mean-variance approach, underscored by a laziness factor. Back-testing of trades showcases the model's superior performance and stability. Future work envisages integrating machine learning for price forecasting and personal trader factors to augment the model's practicality.
	\\	
		
\bibliographystyle{IEEEtran}
\bibliography{IEEEexample}		

\begin{thebibliography}{10}
\providecommand{\url}[1]{#1}
\csname url@samestyle\endcsname
\providecommand{\newblock}{\relax}
\providecommand{\bibinfo}[2]{#2}
\providecommand{\BIBentrySTDinterwordspacing}{\spaceskip=0pt\relax}
\providecommand{\BIBentryALTinterwordstretchfactor}{4}
\providecommand{\BIBentryALTinterwordspacing}{\spaceskip=\fontdimen2\font plus
\BIBentryALTinterwordstretchfactor\fontdimen3\font minus
  \fontdimen4\font\relax}
\providecommand{\BIBforeignlanguage}[2]{{%
\expandafter\ifx\csname l@#1\endcsname\relax
\typeout{** WARNING: IEEEtran.bst: No hyphenation pattern has been}%
\typeout{** loaded for the language `#1'. Using the pattern for}%
\typeout{** the default language instead.}%
\else
\language=\csname l@#1\endcsname
\fi
#2}}
\providecommand{\BIBdecl}{\relax}
\BIBdecl

\bibitem{pan2021roles}
Y.~Pan and L.~Zhang, ``Roles of artificial intelligence in construction
  engineering and management: A critical review and future trends,''
  \emph{Automation in Construction}, vol. 122, p. 103517, 2021.

\bibitem{ashta2021artificial}
A.~Ashta and H.~Herrmann, ``Artificial intelligence and fintech: An overview of
  opportunities and risks for banking, investments, and microfinance,''
  \emph{Strategic Change}, vol.~30, no.~3, pp. 211--222, 2021.

\bibitem{xiao2020stock}
C.~Xiao, W.~Xia, and J.~Jiang, ``Stock price forecast based on combined model
  of ari-ma-ls-svm,'' \emph{Neural Computing and Applications}, vol.~32, pp.
  5379--5388, 2020.

\bibitem{yi2022multifactor}
K.~Yi, ``The multifactor quantitative investment model based on association
  rule mining and machine learning,'' \emph{Wireless Communications and Mobile
  Computing}, vol. 2022, 2022.

\bibitem{alberg2017improving}
J.~Alberg and Z.~C. Lipton, ``Improving factor-based quantitative investing by
  forecasting company fundamentals,'' \emph{arXiv preprint arXiv:1711.04837},
  2017.

\bibitem{leung2015validity}
L.~Leung, ``Validity, reliability, and generalizability in qualitative
  research,'' \emph{Journal of family medicine and primary care}, vol.~4,
  no.~3, p. 324, 2015.

\bibitem{queiros2017strengths}
A.~Queir{\'o}s, D.~Faria, and F.~Almeida, ``Strengths and limitations of
  qualitative and quantitative research methods,'' \emph{European journal of
  education studies}, 2017.

\bibitem{petropoulos2022forecasting}
F.~Petropoulos, D.~Apiletti, V.~Assimakopoulos, M.~Z. Babai, D.~K. Barrow,
  S.~B. Taieb, C.~Bergmeir, R.~J. Bessa, J.~Bijak, J.~E. Boylan \emph{et~al.},
  ``Forecasting: theory and practice,'' \emph{International Journal of
  Forecasting}, 2022.

\bibitem{ta2020portfolio}
V.-D. Ta, C.-M. Liu, and D.~A. Tadesse, ``Portfolio optimization-based stock
  prediction using long-short term memory network in quantitative trading,''
  \emph{Applied Sciences}, vol.~10, no.~2, p. 437, 2020.

\bibitem{binoy2022financial}
S.~J. Binoy and J.~Jos, ``Financial market forecasting using macro-economic
  variables and rnn,'' in \emph{2022 2nd International Conference on Advance
  Computing and Innovative Technologies in Engineering (ICACITE)}.\hskip 1em
  plus 0.5em minus 0.4em\relax IEEE, 2022, pp. 1366--1371.

\bibitem{zhang2022research}
J.~Zhang, Y.~Huang, C.~Huang, W.~Huang \emph{et~al.}, ``Research on arima based
  quantitative investment model,'' \emph{Academic Journal of Business \&
  Management}, vol.~4, no.~17, 2022.

\bibitem{sun2022research}
Y.~Sun, T.~Dong, and B.~Shan, ``Research on stock quantitative investment
  strategy based on arima-garch-mlp model,'' in \emph{2022 IEEE Conference on
  Telecommunications, Optics and Computer Science (TOCS)}.\hskip 1em plus 0.5em
  minus 0.4em\relax IEEE, 2022, pp. 984--990.

\bibitem{han2022quantitative}
N.~Han, S.~Zhang, H.~Wang, Z.~Chen, X.~Hou, and Z.~Sun, ``Quantitative
  investment decision model based on arima and iterative neural network,'' in
  \emph{2022 IEEE Conference on Telecommunications, Optics and Computer Science
  (TOCS)}.\hskip 1em plus 0.5em minus 0.4em\relax IEEE, 2022, pp. 1076--1081.

\bibitem{norouzi2020black}
N.~Norouzi and M.~Fani, ``Black gold falls, black plague arise-an opec crude
  oil price forecast using a gray prediction model,'' \emph{Upstream Oil and
  Gas Technology}, vol.~5, p. 100015, 2020.

\bibitem{de2020using}
O.~V. De~la Torre-Torres, D.~Aguilasocho-Montoya, J.~{\'A}lvarez-Garc{\'\i}a,
  and B.~Simonetti, ``Using markov-switching models with markov chain monte
  carlo inference methods in agricultural commodities trading,'' \emph{Soft
  Computing}, vol.~24, pp. 13\,823--13\,836, 2020.

\bibitem{li2023optimization}
J.~Li, X.~Wang, S.~Ahmad, X.~Huang, and Y.~A. Khan, ``Optimization of
  investment strategies through machine learning,'' \emph{Heliyon}, 2023.

\bibitem{buczynski2021review}
W.~Buczynski, F.~Cuzzolin, and B.~Sahakian, ``A review of machine learning
  experiments in equity investment decision-making: why most published research
  findings do not live up to their promise in real life,'' \emph{International
  Journal of Data Science and Analytics}, vol.~11, pp. 221--242, 2021.

\bibitem{zhang2020research}
K.~Zhang, ``Research on quantitative investment based on machine learning,'' in
  \emph{2020 2nd International Conference on Economic Management and Cultural
  Industry (ICEMCI 2020)}.\hskip 1em plus 0.5em minus 0.4em\relax Atlantis
  Press, 2020, pp. 245--249.

\bibitem{cesari2013trading}
R.~Cesari and M.~Marzo, ``Trading strategies, portfolio monitoring and
  rebalancing,'' \emph{Portfolio Theory and Management, OUP, New York}, pp.
  383--401, 2013.

\bibitem{markowitz2000mean}
H.~M. Markowitz and G.~P. Todd, \emph{Mean-variance analysis in portfolio
  choice and capital markets}.\hskip 1em plus 0.5em minus 0.4em\relax John
  Wiley \& Sons, 2000, vol.~66.

\bibitem{jobson1980estimation}
J.~D. Jobson and B.~Korkie, ``Estimation for markowitz efficient portfolios,''
  \emph{Journal of the American Statistical Association}, vol.~75, no. 371, pp.
  544--554, 1980.

\bibitem{lin2022improved}
J.~Lin, ``Improved markowitz portfolio investment model based on arima model
  and bp neural network,'' in \emph{2022 IEEE 2nd International Conference on
  Data Science and Computer Application (ICDSCA)}.\hskip 1em plus 0.5em minus
  0.4em\relax IEEE, 2022, pp. 504--507.

\bibitem{xidonas2017robust}
P.~Xidonas, G.~Mavrotas, C.~Hassapis, and C.~Zopounidis, ``Robust
  multiobjective portfolio optimization: A minimax regret approach,''
  \emph{European Journal of Operational Research}, vol. 262, no.~1, pp.
  299--305, 2017.

\bibitem{cheng2019risk}
P.~Y. Cheng, ``Risk willingness and perceived utilities to explain risky
  investment choices: a behavioral model,'' \emph{Journal of Behavioral
  Finance}, vol.~20, no.~3, pp. 255--266, 2019.

\bibitem{haensly2020risk}
P.~J. Haensly, ``Risk decomposition, estimation error, and na{\"\i}ve
  diversification,'' \emph{The North American Journal of Economics and
  Finance}, vol.~52, p. 101146, 2020.

\bibitem{warren2019choosing}
G.~J. Warren, ``Choosing and using utility functions in forming portfolios,''
  \emph{Financial Analysts Journal}, vol.~75, no.~3, pp. 39--69, 2019.

\bibitem{du2019implementation}
S.~Du, Y.~Sun, Y.~Hu, and Z.~Lu, ``Implementation of markowitz mean-variance
  model based on matrix-valued factor algorithm,'' in \emph{2019 IEEE 5th Intl
  Conference on Big Data Security on Cloud (BigDataSecurity), IEEE Intl
  Conference on High Performance and Smart Computing,(HPSC) and IEEE Intl
  Conference on Intelligent Data and Security (IDS)}.\hskip 1em plus 0.5em
  minus 0.4em\relax IEEE, 2019, pp. 85--89.

\bibitem{ariyo2014stock}
A.~A. Ariyo, A.~O. Adewumi, and C.~K. Ayo, ``Stock price prediction using the
  arima model,'' in \emph{2014 UKSim-AMSS 16th international conference on
  computer modelling and simulation}.\hskip 1em plus 0.5em minus 0.4em\relax
  IEEE, 2014, pp. 106--112.

\bibitem{lefebvre2020mean}
W.~Lefebvre, G.~Loeper, and H.~Pham, ``Mean-variance portfolio selection with
  tracking error penalization,'' \emph{Mathematics}, vol.~8, no.~11, p. 1915,
  2020.

\end{thebibliography}

\end{document}